# Towards a 'Book Publishers Citation Reports' First approach using the 'Book Citation Index'

**Daniel Torres-Salinas** *
**Nicolás Robinson-García** **
**Emilio Delgado López-Cózar** **

* EC3 Research Group, CIMA, Universidad de Navarra, Pamplona, Spain
** EC3 Research Group, Universidad de Granada, Granada, Spain

**Abstract**: *the absence of books and book chapters in the Web of Science Citation Indexes (SCI, SSCI and A&HCI) has always been considered an important flaw but the Thomson Reuters 'Book Citation Index' database was finally available in October of 2010 indexing 29,618 books and 379,082 book chapters. The Book Citation Index opens a new window of opportunities for analyzing these fields from a bibliometric point of view. The main objective of this article is to analyze different impact indicators referred to the scientific publishers included in the Book Citation Index for the Social Sciences and Humanities fields during 2006-2011. This way we construct what we have called the 'Book Publishers Citation Reports'. For this, we present a total of 19 rankings according to the different disciplines in Humanities & Arts and Social Sciences & Law with six indicators for scientific publishers.*

---

**Corresponding Author**: Daniel Torres Salinas - torressalinas@gmail.com

**Find us on twitter!**: @torressalinas @nrobinson

# Content

Daniel Torres-Salinas, Nicolás Robinson García and Emilio Delgado López-Cózar

# 1. Introduction

The absence of books and book chapters in the *Web of Science Citation Indexes* (SCI, SSCI and A&HCI) has always been considered an important flaw when using this database for bibliometric purposes and especially when assessing fields such as Social Sciences or Humanities in which this publication type plays a major role. In this sense, **Eugene Garfield** as creator of the citation indexes was well aware of this shortcoming and insisted on the necessity of developing a further citation index that would cover this important loophole when stating:

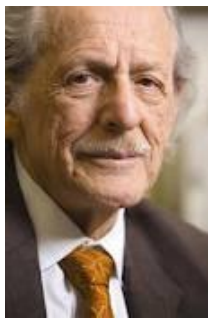

"*From the perspective of the social scientist or humanities scholar, the failure to include monographs as sources in the ISI citation indexes may be a drawback in drawing conclusions about the impact of certain work. Nevertheless, the inclusion of books as cited references in ISI's citation indexes has permitted studies of most-cited books to be accepted as reasonable surrogates for more comprehensive studies that might have included books as sources. Undoubtedly, the creation of a Book Citation Index is a major challenge for the future and would be an expected by-product of the new electronic media with hypertext capability!*"

**Garfield**, 1996

In May 2010 **Thomson Reuters**, intending to put an end to this long criticism, announced at the *Frankfurt Book Fair* the launch of the long-awaited *Book Citation Index* and by the way, getting in ahead of the field. The database was finally available in October of that same year indexing 29,618 books and 379,082 book chapters and covering a time period from 2005 to the present (currently it goes back to 2003) (**Giménez-Toledo** & **Torres-Salinas,** 2011). The emergence of such a product is of great interest not just as an information retrieval tool for Social Sciences and Humanities researchers who finally have an information source to which turn to. But also to bibliometricians and scientific publishers who now have a new tool that includes a long neglected but important publication type such as books which meant a great shortcoming in their studies (**Glänzel** & **Schoepflin**, 1999). The important role books play in Social Sciences and Humanities meant a great threat to any type of approach for research evaluation in these fields as no reliable information source covered them (**Hicks**, 2004) and therefore, were not even considered.

The *Book Citation Index* opens a new window of opportunities for analyzing these fields from a bibliometric point of view (see **Leydesdorff** & **Felt**, 2012). In this sense, the introduction of books in the *Web of Science* platform could lead to some kind of Book Publishers Citation Reports in which scientific publishers would be ranked according to some bibliometric indicator similarly to what the *Journal Citation Reports* does. This would provide another perspective for assessing publishers to those previously presented, for instance analyzing their visibility through their presence in library catalogues (**Torres-Salinas** & **Moed**, 2009) or through surveys to researchers[1]. However, books and journals have different natures that lead to different citation behaviors and must be analyzed and treated carefully. In this line of thought, we present this article in which we pose the possibility of drawing an analogy between scientific publishers and journals. Therefore, our proposal would derive from the traditional journal rankings based on citations. To that effect, the main objective is to analyze different impact indicators referred to the scientific publishers included in the *Book Citation Index* for the Social Sciences and Humanities fields. This way we construct what we have called the '*Book Publishers Citation Reports*'. For this, we present a total of 19 rankings according to the different disciplines in 'Humanities & Arts' and 'Social Sciences & Law' with six indicators for scientific publishers.

---

[1] Scholarly Publisher Indicators: http://epuc.cchs.csic.es/SPI/ranking.html

We believe that databases such as the *Book Citation Index* may lead to the development of new bibliometric tools in order to improve research evaluation exercises. Specially regarding scientific publishers where no tools can be found for measuring objectively and quantitatively their impact within the research community or their level of specialization. In this sense, the '*Book Publishers Citation Reports*' could be hypothetically used similarly in the same way than the current *Journal Citation Reports*, that is, directed to:

- Librarians for facilitating their acquisition process. We must not forget that this was **Eugene Garfield's** original purpose when he created the Journal Impact Factor. These rankings help librarians to differ the core literature in certain disciplines and maximize their budget.

- Researchers for orientating them within the scientific literature. These rankings allow them to rapidly locate which publishers have more visibility and therefore are a good tool when choosing where to send their manuscripts for publication.

- Research managers and bibliometricians as they are powerful tools for research evaluation purposes. In this sense, the '*Book Publishers Citation Reports*' could be used as a proxy for measuring the capability of researchers for instance, to publish in highly cited publishers.

## 2. Methodology

Here we present an analysis of the impact of the scientific publishers included in the *Book Citation Index* for '*Humanities & Arts*' and '*Social Sciences & Law*' in the 2006-2011 time period. We analyzed a total of 19 disciplines which may or may not correspond to the subject categories assigned by the database. In appendix 2 we show the correspondence between the *Web of Science* BKCI subject categories and the disciplines analyzed in this study. In table 1 we present the complete list of disciplines analyzed in this article.

*Table 1* **Set of discipline of Humanities & Arts and Social Sciences & Law selected for the creation of publisher rankings**

| Humanities & Arts | Social Sciences & Law |
|---|---|
| Anthropology | Area & Cultural Studies |
| Archeology | Communication |
| Arts | Economics & Bussiness |
| Geography | Education |
| History | Law |
| History & Philosophy of Science | Information Science & Library Science |
| Language & Linguistics | Managment |
| Literature | Political Science & International Relations |
| Philosophy & Ethics | Sociology |
| Religion | |

Regarding data collecting and processing, in May 2012 the *Book Citation Index* was downloaded and introduced into a relational database where data were processed and indicators calculated. Publishers' names were normalized as many had different variants according to their various headquarters in each country. For instance, for Springer we found variants such as: Springer-Verlag Wien, Springer-Verlag Tokyo, Springer Publishing Co, etc. In table 2 we include the indicators used in our rankings. Regarding the impact indicators, these are the ones obtained at the time of the data download process.

*Table 2* **Set of bibliometric indicators for analyzing the production and impact of publishers included in the Book Citation Index**

| | INDICATOR | ACRONYM | DEFINITION |
|---|---|---|---|
| PRODUCTION | **Number of items indexed** | Total Items | Total records indexed in the Book Citation Index. That is the sum of records indexed as 'book' and 'book chapter'. |
| | **Number of books indexed** | Books | Records indexed as document type 'book' in the Book Citation Index |
| | **Number of book chapters indexed** | Chap | Records indexed as document type 'book chapter' in the Book Citation Index |
| IMPACT | **Total citations received by all items** | Total Citations | Total citations received by all records included in the Book Citation Index. |
| | **Average citations per item** | AvgCit | Average of citations items receive according to their discipline. That is, the result of dividing Total Items between Total Citations. |
| | **Percentage of non cited items** | NonCit | Proportion of items indexed as document type 'book' or 'book chapter' that have received no citations from the total of items of the given discipline. |

# 3. Results

The BKCI has a total of 396,421 records divided in 28,805 books and 367,616 book chapters for the 2006-2011 time period, averaging 12 chapters per book. Considering only the 'Humanities & Arts' and 'Social Sciences & Law fields', they are a total of 17,005 books and 202,830 chapters, averaging 11 chapters per book. This means that Humanities and Social Sciences represent 55% of the total *Book Citation Index*. In table 3 we offer a general perspective of the analyzed disciplines and their production and impact indicators. In this sense, 'Economics & Business', 'Education' and 'History' are the ones with more items indexed and also, and probably as a consequence, the fields with more citations received along with 'Sociology'. On the other side, 'Anthropology' has the highest citation average with 1.68. The non-cited rate ranges from 91% in 'Arts' to 74% in 'Archeology'.

*Table 3* **General overview of the discipline analyzed in this study**

|  | PRODUCTION | | | IMPACT | | |
| --- | --- | --- | --- | --- | --- | --- |
|  | Total Items | Books | Chap | Total Citations | AvgCit | NonCit |
| Anthropology | 3146 | 234 | 2912 | 5280 | 1,68 | 75% |
| Archeology | 2336 | 154 | 2182 | 2367 | 1,01 | 74% |
| Area & Cultural Studies | 15029 | 1273 | 13756 | 7572 | 0,50 | 88% |
| Arts | 1932 | 140 | 1792 | 514 | 0,27 | 91% |
| Communication | 8703 | 596 | 8107 | 4462 | 0,51 | 85% |
| Economics & Bussiness | 35129 | 2577 | 32552 | 24498 | 0,70 | 86% |
| Education | 21068 | 1416 | 19652 | 10360 | 0,49 | 84% |
| Geography | 2670 | 215 | 2455 | 2754 | 1,03 | 79% |
| History | 20346 | 1643 | 18703 | 12067 | 0,59 | 89% |
| History & Philosophy of Science | 5819 | 446 | 5373 | 3081 | 0,53 | 88% |
| Information Science & Library Science | 4235 | 267 | 3968 | 1745 | 0,41 | 85% |
| Languague & Linguistics | 11468 | 760 | 10708 | 7932 | 0,69 | 83% |
| Law | 9824 | 772 | 9052 | 3922 | 0,40 | 88% |
| Literature | 11654 | 1026 | 10628 | 3689 | 0,32 | 90% |
| Managment | 7597 | 543 | 7054 | 4389 | 0,58 | 84% |
| Philosophy & Ethics | 12392 | 944 | 11448 | 6887 | 0,56 | 87% |
| Political Science & Inter. Relations | 31790 | 2750 | 29040 | 26851 | 1,08 | 84% |
| Religion | 8684 | 721 | 7963 | 3795 | 0,44 | 91% |
| Sociology | 9080 | 707 | 8373 | 13464 | 1,48 | 78% |

Finally, in table 2 we show as an example, the bibliometric indicators for the 'Information Science & Library Science' discipline. **The rest of the 18 disciplines are shown in appendix 1**. As observed, all book publishers' rankings per discipline are ordered according to the total number of items per publisher. In the case of 'Information Science & Library Science', the most productive publisher according to the *Book Citation Index* is *Chandos Publishing* (1456 items), followed by *IOS Press* (760 items) and *Springer* (653 items). However, it is worth noting that, while the total items list correlates to a great extent with the number of books (0.9) there are some unexpected results. The most significant is that of *IOS Press* which, according to the *Book Citation Index*, has only 4 books indexed with 756 book chapters, which means an average of 189 chapters per book. The reason for this lies on the fact that this publisher has book series, distorting somehow the picture of publishers' coverage for this discipline. According to the distribution and the concentration of citations and output, this distorting effect of certain publishers affects most of the 19 publishers rankings.

*Table 4* **Output and Impact indicators for publishers in the Information Science & Library Science discipline according to the Book Citation Index**

## Information Science & Library Science

| | PRODUCTION | | | IMPACT | | |
|---|---|---|---|---|---|---|
| | Total Items | Books | Chap | Total Citations | AvgCit | NonCit |
| CHANDOS PUBL | 1456 | 125 | 1331 | 502 | 0,34 | 89% |
| IOS PRESS | 760 | 4 | 756 | 202 | 0,27 | 84% |
| SPRINGER | 653 | 44 | 609 | 353 | 0,54 | 81% |
| WALTER DE GRUYTER & CO | 318 | 18 | 300 | 87 | 0,27 | 88% |
| M E SHARPE INC | 252 | 15 | 237 | 175 | 0,69 | 71% |
| BAYWOOD PUBLISHING CO INC | 154 | 13 | 141 | 34 | 0,22 | 85% |
| EMERALD GROUP PUBLISHING LIMITED | 144 | 13 | 131 | 61 | 0,42 | 75% |
| ROUTLEDGE | 101 | 6 | 95 | 14 | 0,14 | 93% |
| PALGRAVE | 100 | 4 | 96 | 7 | 0,07 | 96% |
| M I T PRESS | 47 | 4 | 43 | 34 | 0,72 | 87% |
| WOODHEAD PUBL LTD | 41 | 4 | 37 | 10 | 0,24 | 90% |
| NOVA SCIENCE PUBLISHERS, INC | 28 | 3 | 25 | 0 | 0,00 | 100% |
| CAMBRIDGE UNIV PRESS | 26 | 2 | 24 | 18 | 0,69 | 92% |
| TMC ASSER PRESS | 26 | 1 | 25 | 0 | 0,00 | 100% |
| ELSEVIER | 25 | 2 | 23 | 128 | 5,12 | 92% |
| EDWARD ELGAR PUBLISHING LTD | 23 | 2 | 21 | 31 | 1,35 | 91% |
| CABI PUBLISHING-C A B INT | 21 | 1 | 20 | 50 | 2,38 | 48% |
| WORLD SCIENTIFIC PUBL CO PTE LTD | 18 | 1 | 17 | 8 | 0,44 | 89% |
| UNIV ADELAIDE PRESS | 9 | 1 | 8 | 0 | 0,00 | 100% |
| UTAH STATE UNIV PRESS | 9 | 1 | 8 | 1 | 0,11 | 89% |
| CRC PRESS-TAYLOR & FRANCIS GROUP | 8 | 1 | 7 | 0 | 0,00 | 100% |
| UNIV CALIFORNIA PRESS | 8 | 1 | 7 | 27 | 3,38 | 75% |
| WILFRID LAURIER UNIV PRESS | 8 | 1 | 7 | 3 | 0,38 | 75% |

# 4. Discussion and concluding remarks

*Discussion*

If we observe the list of rankings presented in appendix 1, we can conclude that it is possible to develop a so-called '*Book Publishers Citation Reports*' based on the *Book Citation Index (Thomson Reuters)*. That is, when taking into account the technical issues that must be surpassed to do so, however, we must emphasize different problems we have encountered that warns us against the use of such a tool for evaluating purposes. The results offered by the *Book Citation Index* could not be valid or reliable for bibliometric use, although they may be a good tool for academic librarians. The issues we have encountered can be resumed in the following way:

- There is a clear dominance of English-language publishers with a commercial profile. When observed, practically all rankings are led by international commercial publishers such as *Springer*, *Routledge* or *Palgrave*. Poor presence of university presses. Only those from *Princeton*, Cambridge, *California* or the *Australian National University* have a notable presence.

- There is almost no representation of those countries with an important scientific background in the Humanities and Social Sciences such as Italy, France or Germany. In fact, France for instance has no publishers indexed, neglecting for example *Editoriales Presses Universitaires de France (PUF)*. In the case of Spain, publishers such as *Ariel* or *Alianza* for example, which are greatly considered by Spanish researchers as shown in the *Scholarly Publishers Indicators project*[2] are omitted. Therefore, they have not even considered introducing the most important publishers per region or country.

- There is a surprising absence, limited presence or underrepresentation of globally important publishers such as *Peter Lang*, *Pearson*, and *Macmillan* or of specialized publishers such as *John Benjamins* for Linguistics, *Giuffrè* for Law or *Archaeopress* for Archeology.

These problems are especially severe in the case of Humanities and Social Sciences where there is no such a thing as a global scientific community as it happens in Basic and Applied Sciences, and where English is not considered as the main scientific language. We must take into account the effect of the local and national factor and the atomization of knowledge that exist in these areas. According to *Thomson Reuters*' official report they only consider English-language publications 'Because English is the universal language of science at this time, Thomson Reuters will focus on books that publish full text in English[3']. In our opinion, this is an unfortunate statement when regarding to these fields.

---

[2] http://epuc.cchs.csic.es/SPI/index.html

Finally, we must point out several issues when developing publishers' indicators:

1) What must we count, books or book chapters? must we add their citations? Should we count book citations and chapters citations separately? should we distinguish between multi-authored books or single-authored book?

2) What should we do with those monographs which behave more closely to journals than the rest such as book series as *Annual Reviews*? Should they be excluded in order to end with their distorting effect?

3) Although this has not been analyzed in this study, which is the most suitable citation window for measuring books' impact? Can we preserve the Journal Impact Factor analogy?

*Concluding remarks*

►Thanks to the launch of the *Book Citation Index*, it is currently possible to develop scientific publisher's rankings similar to the Thomson Reuters '*Journal Citation Reports*'. We called these rankings '*Book Publishers Citation Reports*'.

► The 19 rankings presented in this paper could be only useful for characterize the Anglosaxon domain because there is a strong bias to commercial and English speaking countries. So is not possible to develop a global '*Book Publishers Citation Reports*'.

►There is a lack of important publishers so we obtain a partial picture of the publisher's impact scenario. A '*Book Publishers Citation Report*' based in the BKCI maybe is no useful for research evaluation purposes in the same way as the JCR is.

---

[3] http://wokinfo.com/media/pdf/BKCI-SelectionEssay_web.pdf

Torres-Salinas, D., & Moed, H.F. (2009). *Library Catalog Analysis as a tool in studies of social sciences and humanities: An exploratory study of published book titles in Economics*. Journal of Informetrics, 3(1), 9-26.

## Appendix 1. Publisher rankings

### Anthropology

| | PRODUCTION | | | IMPACT | | |
|---|---|---|---|---|---|---|
| | Total Items | Books | Chap | Total Citations | AvgCit | NonCit |
| UNIV CALIFORNIA PRESS | 748 | 61 | 687 | 1172 | 1,57 | 85% |
| SPRINGER | 534 | 32 | 502 | 611 | 1,14 | 63% |
| ROUTLEDGE | 367 | 20 | 347 | 139 | 0,38 | 82% |
| PALGRAVE | 248 | 22 | 226 | 168 | 0,68 | 93% |
| AUSTRALIAN NATL UNIV | 189 | 13 | 176 | 56 | 0,30 | 81% |
| UNIV PENNSYLVANIA PRESS | 169 | 18 | 151 | 208 | 1,23 | 76% |
| ANNUAL REVIEWS | 166 | 6 | 160 | 1853 | 11,16 | 14% |
| CAMBRIDGE UNIV PRESS | 145 | 12 | 133 | 192 | 1,32 | 74% |
| PRINCETON UNIV PRESS | 121 | 10 | 111 | 375 | 3,10 | 92% |
| UNIV PRESS COLORADO | 113 | 8 | 105 | 69 | 0,61 | 77% |
| EMERALD GROUP | 66 | 5 | 61 | 51 | 0,77 | 65% |
| WILEY-LISS, INC | 47 | 6 | 41 | 330 | 7,02 | 30% |
| E J BRILL | 38 | 4 | 34 | 3 | 0,08 | 95% |
| AMER ANTHROPOLOGICAL ASSOC | 36 | 3 | 33 | 25 | 0,69 | 64% |
| UTAH STATE UNIV PRESS | 29 | 2 | 27 | 1 | 0,03 | 97% |
| ATHABASCA UNIV PRESS | 28 | 2 | 26 | 8 | 0,29 | 96% |
| NOVA SCIENCE PUBLISHERS, INC | 19 | 2 | 17 | 0 | 0,00 | 100% |
| HUMANA PRESS INC | 13 | 1 | 12 | 1 | 0,08 | 92% |
| INTELLECT LTD | 13 | 1 | 12 | 10 | 0,77 | 77% |
| WALTER DE GRUYTER & CO | 12 | 1 | 11 | 5 | 0,42 | 58% |
| UNIV WEST BOHEMIA | 11 | 1 | 10 | 0 | 0,00 | 100% |
| UNIVERSITY ALASKA PRESS | 11 | 1 | 10 | 0 | 0,00 | 100% |
| WILFRID LAURIER UNIV PRESS | 11 | 1 | 10 | 0 | 0,00 | 100% |
| PRESSES UNIV MIRAIL | 10 | 1 | 9 | 2 | 0,20 | 80% |
| UNIV NORTH CAROLINA PRESS | 1 | | 1 | 0 | 0,00 | 100% |
| UNIV WASHINGTON PRESS | 1 | 1 | | 1 | 1,00 | 0% |

Daniel Torres-Salinas, Nicolás Robinson García and Emilio Delgado López-Cózar

## Archeology

| | PRODUCTION | | | IMPACT | | |
|---|---|---|---|---|---|---|
| | Total Items | Books | Chap | Total Citations | AvgCit | NonCit |
| SPRINGER | 878 | 57 | 821 | 1066 | 1,21 | 63% |
| ROUTLEDGE | 280 | 9 | 271 | 58 | 0,21 | 89% |
| WALTER DE GRUYTER & CO | 261 | 21 | 240 | 24 | 0,09 | 94% |
| AUSTRALIAN NATL UNIV | 238 | 13 | 225 | 220 | 0,92 | 68% |
| CAMBRIDGE UNIV PRESS | 187 | 16 | 171 | 370 | 1,98 | 84% |
| UNIV CALIFORNIA PRESS | 73 | 6 | 67 | 149 | 2,04 | 85% |
| PRINCETON UNIV PRESS | 44 | 3 | 41 | 76 | 1,73 | 93% |
| UNIV PENNSYLVANIA PRESS | 42 | 4 | 38 | 106 | 2,52 | 55% |
| UNIV PRESS COLORADO | 38 | 3 | 35 | 26 | 0,68 | 68% |
| AMER ANTHROPOLOGICAL ASSOC | 36 | 3 | 33 | 25 | 0,69 | 64% |
| NOVA SCIENCE PUBLISHERS, INC | 30 | 3 | 27 | 6 | 0,20 | 93% |
| EUROPAE ARCH.CONSILIUM-EAC | 25 | 1 | 24 | 1 | 0,04 | 96% |
| CRC PRESS-TAYLOR & FRANCIS | 24 | 1 | 23 | 31 | 1,29 | 54% |
| UNIV BRITISH COLUMBIA PRESS | 24 | 1 | 23 | 57 | 2,38 | 38% |
| TAYLOR & FRANCIS LTD | 20 | 2 | 18 | 1 | 0,05 | 95% |
| ROYAL SOC CHEMISTRY | 18 | 1 | 17 | 20 | 1,11 | 94% |
| UNIV PENNSYLVANIA MUSEUM | 18 | 1 | 17 | 36 | 2,00 | 17% |
| DAS ES SALAAM UNIV PRESS LTD | 15 | 1 | 14 | 3 | 0,20 | 80% |
| M I T PRESS | 14 | 1 | 13 | 17 | 1,21 | 86% |
| PALGRAVE | 14 | 1 | 13 | 6 | 0,43 | 79% |
| HUMANA PRESS INC | 13 | 1 | 12 | 1 | 0,08 | 92% |
| NOTTINGHAM UNIVERSITY PRESS | 12 | 1 | 11 | 1 | 0,08 | 92% |
| UNIV WEST BOHEMIA | 11 | 1 | 10 | 0 | 0,00 | 100% |
| ATHABASCA UNIV PRESS | 10 | 1 | 9 | 2 | 0,20 | 90% |
| BLACKWELL SCIENCE PUBL | 10 | 1 | 9 | 65 | 6,50 | 30% |
| EDITURA EXCELSIOR ART | 1 | 1 | | 0 | 0,00 | 100% |

# Area & Cultural Studies

| | PRODUCTION | | | IMPACT | | |
|---|---|---|---|---|---|---|
| | Total Items | Books | Chap | Total Citations | AvgCit | NonCit |
| ROUTLEDGE | 6740 | 578 | 6162 | 2890 | 0,43 | 88% |
| PALGRAVE | 3562 | 302 | 3260 | 1405 | 0,39 | 87% |
| UNIV PENNSYLVANIA PRESS | 622 | 66 | 556 | 552 | 0,89 | 81% |
| SPRINGER | 578 | 43 | 535 | 330 | 0,57 | 78% |
| UNIV CALIFORNIA PRESS | 510 | 45 | 465 | 563 | 1,10 | 89% |
| CHANDOS PUBL | 386 | 32 | 354 | 108 | 0,28 | 92% |
| PRINCETON UNIV PRESS | 298 | 26 | 272 | 828 | 2,78 | 89% |
| NOVA SCIENCE PUBLISHERS, INC | 279 | 28 | 251 | 20 | 0,07 | 94% |
| EDWARD ELGAR PUBLISHING LTD | 213 | 16 | 197 | 53 | 0,25 | 92% |
| UNIV NORTH CAROLINA PRESS | 171 | 14 | 157 | 145 | 0,85 | 90% |
| CAMBRIDGE UNIV PRESS | 164 | 15 | 149 | 216 | 1,32 | 90% |
| WILFRID LAURIER UNIV PRESS | 156 | 8 | 148 | 34 | 0,22 | 85% |
| EDITIONS RODOPI B V | 145 | 6 | 139 | 3 | 0,02 | 98% |
| E J BRILL | 144 | 12 | 132 | 9 | 0,06 | 96% |
| US INST PEACE | 119 | 9 | 110 | 90 | 0,76 | 75% |
| AUSTRALIAN NATL UNIV | 111 | 8 | 103 | 8 | 0,07 | 96% |
| INST SOUTHEAST ASIAN STUDIES | 101 | 9 | 92 | 5 | 0,05 | 95% |
| ATHABASCA UNIV PRESS | 83 | 6 | 77 | 9 | 0,11 | 94% |
| BULLETIN ASIA INST | 77 | 4 | 73 | 8 | 0,10 | 90% |
| INFORMATION AGE PUBLISHING-IAP | 73 | 5 | 68 | 13 | 0,18 | 90% |
| M I T PRESS | 59 | 5 | 54 | 122 | 2,07 | 85% |
| PURDUE UNIV PRESS | 59 | 6 | 53 | 19 | 0,32 | 86% |
| UNIV BRITISH COLUMBIA PRESS | 45 | 4 | 41 | 6 | 0,13 | 96% |
| WORLD SCIENTIFIC PUBL CO PTE LTD | 34 | 1 | 33 | 1 | 0,03 | 97% |
| UNIV MINNESOTA PRESS | 33 | 4 | 29 | 16 | 0,48 | 91% |
| CENTRAL EUROPEAN UNIV PRESS | 31 | 1 | 30 | 4 | 0,13 | 97% |
| UNIV PRESS KENTUCKY | 27 | 2 | 25 | 4 | 0,15 | 96% |
| GALLAUDET UNIV PR | 23 | 1 | 22 | 0 | 0,00 | 100% |
| WALTER DE GRUYTER & CO | 23 | 2 | 21 | 0 | 0,00 | 100% |
| EDINBURGH UNIV PRESS | 21 | 2 | 19 | 0 | 0,00 | 100% |
| TAYLOR & FRANCIS LTD | 18 | 1 | 17 | 3 | 0,17 | 89% |
| INTELLECT LTD | 17 | 2 | 15 | 3 | 0,18 | 88% |
| CSIRO PUBLISHING | 14 | 1 | 13 | 0 | 0,00 | 100% |
| PSYCHOLOGY PRESS | 14 | 1 | 13 | 10 | 0,71 | 93% |
| OHIO UNIV PRESS | 13 | 1 | 12 | 17 | 1,31 | 92% |
| BLACKWELL PUBL | 12 | 1 | 11 | 0 | 0,00 | 100% |
| FORDHAM UNIV PRESS | 11 | 1 | 10 | 1 | 0,09 | 91% |
| WORLD BANK INST | 10 | 1 | 9 | 60 | 6,00 | 90% |
| UTAH STATE UNIV PRESS | 9 | 1 | 8 | 0 | 0,00 | 100% |
| MANEY PUBLISHING | 8 | 1 | 7 | 1 | 0,13 | 88% |
| WELLCOME INST HISTORY MEDICINE | 8 | 1 | 7 | 1 | 0,13 | 88% |
| MONASH UNIV PUBLISHING | 4 | | 4 | 1 | 0,25 | 75% |
| WORLD HEALTH ORGANIZATION | 4 | 1 | 3 | 14 | 3,50 | 75% |

Daniel Torres-Salinas, Nicolás Robinson García and Emilio Delgado López-Cózar

## Arts

| | PRODUCTION | | | IMPACT | | |
|---|---|---|---|---|---|---|
| | Total Items | Books | Chap | Total Citations | AvgCit | NonCit |
| ROUTLEDGE | 374 | 22 | 352 | 76 | 0,20 | 93% |
| INTELLECT LTD | 316 | 22 | 294 | 67 | 0,21 | 91% |
| PALGRAVE | 266 | 19 | 247 | 79 | 0,30 | 89% |
| SPRINGER | 146 | 7 | 139 | 16 | 0,11 | 94% |
| UNIV PENNSYLVANIA PRESS | 129 | 14 | 115 | 54 | 0,42 | 91% |
| UNIV NORTH CAROLINA PRESS | 81 | 5 | 76 | 27 | 0,33 | 94% |
| UNIV CALIFORNIA PRESS | 73 | 7 | 66 | 29 | 0,40 | 90% |
| MANEY PUBLISHING | 61 | 4 | 57 | 13 | 0,21 | 90% |
| EDITIONS RODOPI B V | 56 | 3 | 53 | 9 | 0,16 | 88% |
| BIRKHAUSER VERLAG AG | 54 | 3 | 51 | 1 | 0,02 | 98% |
| CAMBRIDGE UNIV PRESS | 53 | 5 | 48 | 28 | 0,53 | 89% |
| E J BRILL | 46 | 3 | 43 | 4 | 0,09 | 91% |
| UITGEVERIJ WAANDERS BV | 40 | 3 | 37 | 1 | 0,03 | 98% |
| AUSTRALIAN NATL UNIV | 29 | 2 | 27 | 1 | 0,03 | 97% |
| M I T PRESS | 29 | 2 | 27 | 10 | 0,34 | 90% |
| PRINCETON UNIV PRESS | 28 | 3 | 25 | 37 | 1,32 | 82% |
| UNIV MINNESOTA PRESS | 26 | 3 | 23 | 4 | 0,15 | 88% |
| ARTS EXTENSION SERV | 16 | 1 | 15 | 2 | 0,13 | 94% |
| WALTER DE GRUYTER & CO | 16 | 2 | 14 | 4 | 0,25 | 94% |
| ROYAL SOC CHEMISTRY | 14 | 1 | 13 | 16 | 1,14 | 57% |
| EDWARD ELGAR PUBLISHING LTD | 13 | 1 | 12 | 21 | 1,62 | 69% |
| MONASH UNIV PUBLISHING | 12 | 1 | 11 | 0 | 0,00 | 100% |
| BLACKWELL SCIENCE PUBL | 11 | 1 | 10 | 14 | 1,27 | 91% |
| TAYLOR & FRANCIS LTD | 11 | 1 | 10 | 0 | 0,00 | 100% |
| LARS MULLER PUBL | 10 | 1 | 9 | 0 | 0,00 | 100% |
| HUMANA PRESS INC | 9 | 1 | 8 | 1 | 0,11 | 89% |
| UNIV WASHINGTON PRESS | 8 | 1 | 7 | 0 | 0,00 | 100% |
| WILFRID LAURIER UNIV PRESS | 4 | 1 | 3 | 0 | 0,00 | 100% |
| VILNIUS GEDIMINAS TECHNICAL UP | 1 | 1 | | 0 | 0,00 | 100% |

# Communication

| | PRODUCTION | | | IMPACT | | |
|---|---|---|---|---|---|---|
| | Total Items | Books | Chap | Total Citations | AvgCit | NonCit |
| ROUTLEDGE | 2490 | 164 | 2326 | 1356 | 0,54 | 84% |
| PALGRAVE | 2078 | 168 | 1910 | 889 | 0,43 | 87% |
| WALTER DE GRUYTER & CO | 1020 | 30 | 990 | 287 | 0,28 | 88% |
| INTELLECT LTD | 841 | 63 | 778 | 353 | 0,42 | 85% |
| UNIV CALIFORNIA PRESS | 503 | 37 | 466 | 300 | 0,60 | 89% |
| EDINBURGH UNIV PRESS | 189 | 17 | 172 | 63 | 0,33 | 93% |
| SPRINGER | 184 | 11 | 173 | 61 | 0,33 | 89% |
| BAYWOOD PUBLISHING CO INC | 154 | 13 | 141 | 34 | 0,22 | 85% |
| M I T PRESS | 101 | 7 | 94 | 435 | 4,31 | 46% |
| CAMBRIDGE UNIV PRESS | 96 | 6 | 90 | 40 | 0,42 | 88% |
| CHANDOS PUBL | 95 | 8 | 87 | 45 | 0,47 | 80% |
| TAYLOR & FRANCIS LTD | 86 | 5 | 81 | 4 | 0,05 | 95% |
| PURDUE UNIV PRESS | 85 | 4 | 81 | 3 | 0,04 | 98% |
| WILFRID LAURIER UNIV PRESS | 85 | 6 | 79 | 18 | 0,21 | 86% |
| PRINCETON UNIV PRESS | 78 | 8 | 70 | 156 | 2,00 | 86% |
| UTAH STATE UNIV PRESS | 75 | 6 | 69 | 42 | 0,56 | 87% |
| LAWRENCE ERLBAUM ASSOC PUBL | 74 | 5 | 69 | 147 | 1,99 | 39% |
| UNIV MINNESOTA PRESS | 63 | 6 | 57 | 7 | 0,11 | 94% |
| NOVA SCIENCE PUBLISHERS, INC | 58 | 6 | 52 | 0 | 0,00 | 100% |
| IOS PRESS | 57 | 3 | 54 | 77 | 1,35 | 70% |
| INFORMATION AGE PUBLISHING-IAP | 49 | 4 | 45 | 5 | 0,10 | 94% |
| EDWARD ELGAR PUBLISHING LTD | 36 | 3 | 33 | 8 | 0,22 | 86% |
| BIRKHAUSER VERLAG AG | 33 | 1 | 32 | 0 | 0,00 | 100% |
| UNIV NORTH CAROLINA PRESS | 29 | 1 | 28 | 4 | 0,14 | 97% |
| UNIV PENNSYLVANIA PRESS | 25 | 3 | 22 | 29 | 1,16 | 84% |
| CABI PUBLISHING-C A B INT | 18 | 1 | 17 | 8 | 0,44 | 72% |
| MONASH UNIV PUBLISHING | 18 | 1 | 17 | 4 | 0,22 | 89% |
| AUSTRALIAN NATL UNIV | 13 | 1 | 12 | 10 | 0,77 | 62% |
| ROWMAN & LITTLEFIELD PUBL INC | 13 | 1 | 12 | 74 | 5,69 | 23% |
| UNIV PRESS KENTUCKY | 13 | 1 | 12 | 3 | 0,23 | 92% |
| E J BRILL | 12 | 1 | 11 | 0 | 0,00 | 100% |
| FORDHAM UNIV PRESS | 12 | 1 | 11 | 0 | 0,00 | 100% |
| ATHABASCA UNIV PRESS | 9 | 1 | 8 | 0 | 0,00 | 100% |
| UNIV BRITISH COLUMBIA PRESS | 9 | 1 | 8 | 0 | 0,00 | 100% |
| EDITURA CU/ART | 1 | 1 | | 0 | 0,00 | 100% |
| US INST PEACE | 1 | 1 | | 0 | 0,00 | 100% |

Daniel Torres-Salinas, Nicolás Robinson García and Emilio Delgado López-Cózar

# Economics & Bussiness

| | PRODUCTION | | | IMPACT | | |
|---|---|---|---|---|---|---|
| | Total Items | Books | Chap | Total Citations | AvgCit | NonCit |
| SPRINGER | 7282 | 575 | 6707 | 5586 | 0,77 | 86% |
| PALGRAVE | 6839 | 538 | 6301 | 2522 | 0,37 | 88% |
| ROUTLEDGE | 5944 | 431 | 5513 | 1921 | 0,32 | 89% |
| EDWARD ELGAR PUBLISHING LTD | 5503 | 372 | 5131 | 2685 | 0,49 | 85% |
| PRINCETON UNIV PRESS | 1748 | 132 | 1616 | 6044 | 3,46 | 87% |
| EMERALD GROUP | 1480 | 110 | 1370 | 824 | 0,56 | 74% |
| IOS PRESS | 918 | 9 | 909 | 235 | 0,26 | 85% |
| CAMBRIDGE UNIV PRESS | 787 | 55 | 732 | 2026 | 2,57 | 68% |
| NOVA SCIENCE PUBLISHERS, INC | 770 | 75 | 695 | 71 | 0,09 | 93% |
| PHYSICA-VERLAG GMBH & CO | 486 | 43 | 443 | 171 | 0,35 | 84% |
| M I T PRESS | 389 | 27 | 362 | 585 | 1,50 | 75% |
| AUSTRALIAN NATL UNIV | 349 | 24 | 325 | 222 | 0,64 | 80% |
| CHANDOS PUBL | 194 | 15 | 179 | 45 | 0,23 | 94% |
| INFORMATION AGE PUBLISHING-IAP | 185 | 14 | 171 | 68 | 0,37 | 85% |
| UNIV PENNSYLVANIA PRESS | 179 | 15 | 164 | 164 | 0,92 | 74% |
| ANNUAL REVIEWS | 158 | 7 | 151 | 424 | 2,68 | 34% |
| RC PRESS-TAYLOR & FRANCIS GROUP | 151 | 10 | 141 | 37 | 0,25 | 95% |
| UNIV CALIFORNIA PRESS | 141 | 12 | 129 | 144 | 1,02 | 93% |
| CHAPMAN & HALL/CRC PRESS | 130 | 7 | 123 | 36 | 0,28 | 85% |
| WALTER DE GRUYTER & CO | 123 | 9 | 114 | 7 | 0,06 | 97% |
| WORLD BANK INST | 121 | 10 | 111 | 174 | 1,44 | 74% |
| JAI PRESS INC | 100 | 6 | 94 | 67 | 0,67 | 83% |
| WORLD SCIENTIFIC PUBL CO PTE LTD | 84 | 7 | 77 | 14 | 0,17 | 89% |
| WILFRID LAURIER UNIV PRESS | 79 | 3 | 76 | 13 | 0,16 | 92% |
| TAYLOR & FRANCIS LTD | 74 | 4 | 70 | 3 | 0,04 | 97% |
| UNIV ADELAIDE PRESS | 74 | 3 | 71 | 0 | 0,00 | 100% |
| IGI GLOBAL | 73 | 4 | 69 | 12 | 0,16 | 93% |
| UNIV NORTH CAROLINA PRESS | 68 | 5 | 63 | 12 | 0,18 | 94% |
| AUERBACH PUBLICATIONS | 66 | 4 | 62 | 12 | 0,18 | 94% |
| BLACKWELL SCIENCE PUBL | 53 | 2 | 51 | 0 | 0,00 | 100% |
| JAI-ELSEVIER LTD | 46 | 3 | 43 | 12 | 0,26 | 80% |
| CABI PUBLISHING-C A B INT | 45 | 2 | 43 | 162 | 3,60 | 22% |
| JAI-ELSEVIER SCIENCE INC | 44 | 5 | 39 | 1 | 0,02 | 98% |
| UNIV CHICAGO PRESS | 34 | 2 | 32 | 41 | 1,21 | 59% |
| UNIV BRITISH COLUMBIA PRESS | 32 | 3 | 29 | 18 | 0,56 | 91% |
| EDINBURGH UNIV PRESS | 31 | 3 | 28 | 1 | 0,03 | 97% |
| ELSEVIER | 31 | 2 | 29 | 13 | 0,42 | 81% |
| INST SOUTHEAST ASIAN STUDIES | 29 | 2 | 27 | 4 | 0,14 | 86% |
| APRESS | 28 | 2 | 26 | 0 | 0,00 | 100% |
| INDEPENDENT INST | 22 | 1 | 21 | 5 | 0,23 | 86% |
| TRANSACTION PUBLISHERS | 21 | 1 | 20 | 27 | 1,29 | 38% |
| PENNSYLVANIA STATE UNIV PRESS | 20 | 1 | 19 | 11 | 0,55 | 95% |
| STANFORD UNIV PRESS | 20 | 1 | 19 | 1 | 0,05 | 95% |
| ATHABASCA UNIV PRESS | 18 | 2 | 16 | 1 | 0,06 | 94% |
| BROOKINGS INST | 17 | 2 | 15 | 4 | 0,24 | 88% |
| CA PRESS | 13 | 1 | 12 | 0 | 0,00 | 100% |
| LARS MULLER PUBL | 13 | 1 | 12 | 0 | 0,00 | 100% |
| MARTIN STRIZ PUBLISHING | 13 | 1 | 12 | 0 | 0,00 | 100% |
| PURDUE UNIV PRESS | 13 | 1 | 12 | 0 | 0,00 | 100% |
| SPON PRESS | 13 | 1 | 12 | 0 | 0,00 | 100% |
| INTELLECT LTD | 12 | 1 | 11 | 4 | 0,33 | 92% |
| PSYCHOLOGY PRESS | 10 | 1 | 9 | 13 | 1,30 | 90% |
| E J BRILL | 9 | 1 | 8 | 2 | 0,22 | 89% |
| PICKERING & CHATTO PUBLISHERS | 9 | 1 | 8 | 0 | 0,00 | 100% |
| WOODHEAD PUBL LTD | 9 | 1 | 8 | 0 | 0,00 | 100% |
| EAST-WEST CENTER | 8 | 1 | 7 | 2 | 0,25 | 88% |
| UTAH STATE UNIV PRESS | 8 | 1 | 7 | 9 | 1,13 | 88% |
| INT MONETARY FUND | 5 | 1 | 4 | 39 | 7,80 | 80% |
| VYDAVATELSTVI A… | 5 | 1 | 4 | 0 | 0,00 | 100% |
| US INST PEACE | 2 | 2 | | 3 | 1,50 | 0% |
| UNIV NOTRE DAME PRESS | 1 | 1 | | 1 | 1,00 | 0% |

## Education

| | PRODUCTION | | | IMPACT | | |
|---|---|---|---|---|---|---|
| | Total Items | Books | Chap | Total Citations | AvgCit | NonCit |
| SPRINGER | 7112 | 428 | 6684 | 3812 | 0,54 | 80% |
| ROUTLEDGE | 4231 | 280 | 3951 | 1807 | 0,43 | 86% |
| INFORMATION AGE PUBLISHING-IAP | 2944 | 214 | 2730 | 918 | 0,31 | 88% |
| PALGRAVE | 2637 | 206 | 2431 | 1199 | 0,45 | 84% |
| NOVA SCIENCE PUBLISHERS, INC | 573 | 52 | 521 | 28 | 0,05 | 96% |
| EMERALD GROUP | 445 | 30 | 415 | 199 | 0,45 | 74% |
| UTAH STATE UNIV PRESS | 344 | 24 | 320 | 136 | 0,40 | 87% |
| PRINCETON UNIV PRESS | 254 | 21 | 233 | 442 | 1,74 | 92% |
| TAYLOR & FRANCIS LTD | 241 | 11 | 230 | 5 | 0,02 | 98% |
| ELSEVIER | 190 | 15 | 175 | 237 | 1,25 | 52% |
| EDWARD ELGAR PUBLISHING LTD | 179 | 14 | 165 | 113 | 0,63 | 87% |
| INTELLECT LTD | 150 | 8 | 142 | 20 | 0,13 | 95% |
| WILEY-BLACKWELL | 130 | 5 | 125 | 60 | 0,46 | 81% |
| CLAME | 129 | 1 | 128 | 0 | 0,00 | 100% |
| WALTER DE GRUYTER & CO | 124 | 10 | 114 | 72 | 0,58 | 77% |
| CAMBRIDGE UNIV PRESS | 123 | 9 | 114 | 264 | 2,15 | 74% |
| BLACKWELL SCIENCE PUBL | 101 | 3 | 98 | 35 | 0,35 | 77% |
| UNIV CALIFORNIA PRESS | 91 | 6 | 85 | 39 | 0,43 | 93% |
| NSTA PRESS | 79 | 7 | 72 | 11 | 0,14 | 91% |
| ATHABASCA UNIV PRESS | 73 | 4 | 69 | 57 | 0,78 | 67% |
| GALLAUDET UNIV PR | 71 | 6 | 65 | 20 | 0,28 | 83% |
| LAWRENCE ERLBAUM ASSOC PUBL | 60 | 5 | 55 | 77 | 1,28 | 83% |
| WILFRID LAURIER UNIV PRESS | 52 | 2 | 50 | 15 | 0,29 | 87% |
| PSYCHOLOGY PRESS | 51 | 3 | 48 | 113 | 2,22 | 57% |
| PURDUE UNIV PRESS | 47 | 2 | 45 | 4 | 0,09 | 96% |
| SAGE PUBLICATIONS INC | 47 | 4 | 43 | 293 | 6,23 | 15% |
| CHANDOS PUBL | 42 | 4 | 38 | 7 | 0,17 | 88% |
| UNIV NORTH CAROLINA PRESS | 40 | 3 | 37 | 35 | 0,88 | 90% |
| UNIV BRITISH COLUMBIA PRESS | 39 | 3 | 36 | 4 | 0,10 | 95% |
| EDINBURGH UNIV PRESS | 37 | 3 | 34 | 0 | 0,00 | 100% |
| WAGENINGEN ACADEMIC PUBLISHERS | 32 | 1 | 31 | 0 | 0,00 | 100% |
| JAI PRESS INC | 30 | 2 | 28 | 9 | 0,30 | 77% |
| AUSTRALIAN NATL UNIV | 28 | 2 | 26 | 9 | 0,32 | 79% |
| IOS PRESS | 27 | 2 | 25 | 4 | 0,15 | 85% |
| FORDHAM UNIV PRESS | 25 | 2 | 23 | 0 | 0,00 | 100% |
| IGI GLOBAL | 24 | 1 | 23 | 2 | 0,08 | 92% |
| MATHEMATICAL ASSOC AMERICA | 23 | 1 | 22 | 16 | 0,70 | 70% |
| JAI-ELSEVIER SCIENCE INC | 22 | 2 | 20 | 0 | 0,00 | 100% |
| SYMPOSIUM BOOKS | 22 | 2 | 20 | 9 | 0,41 | 73% |
| IDEA GROUP PUBLISHING | 21 | 1 | 20 | 65 | 3,10 | 43% |
| MONASH UNIV PUBLISHING | 20 | 1 | 19 | 33 | 1,65 | 55% |
| AMERICAN EDUCATIONAL RESEARCH ASS. | 19 | 2 | 17 | 113 | 5,95 | 21% |
| UNIV PENNSYLVANIA PRESS | 17 | 3 | 14 | 55 | 3,24 | 82% |
| NOTTINGHAM UNIVERSITY PRESS | 15 | 1 | 14 | 1 | 0,07 | 93% |
| UNIV PRESS MISSISSIPPI | 14 | 1 | 13 | 0 | 0,00 | 100% |
| OXFORD UNIV PRESS | 13 | 1 | 12 | 0 | 0,00 | 100% |
| ROYAL SOC CHEMISTRY | 12 | 1 | 11 | 4 | 0,33 | 75% |
| CRC PRESS-TAYLOR & FRANCIS | 11 | 1 | 10 | 1 | 0,09 | 91% |
| M I T PRESS | 11 | 1 | 10 | 4 | 0,36 | 91% |
| PHYSICA-VERLAG GMBH & CO | 11 | 1 | 10 | 3 | 0,27 | 91% |
| UNIV MINNESOTA PRESS | 11 | 1 | 10 | 1 | 0,09 | 91% |
| INST SOUTHEAST ASIAN STUDIES | 9 | 1 | 8 | 0 | 0,00 | 100% |
| CABI PUBLISHING-C A B INT | 8 | 1 | 7 | 9 | 1,13 | 88% |
| VS VERLAG SOZIALWISSENSCHAFTEN | 7 | 1 | 6 | 0 | 0,00 | 100% |

## History

| | PRODUCTION | | | IMPACT | | |
|---|---|---|---|---|---|---|
| | Total Items | Books | Chap | Total Citations | AvgCit | NonCit |
| PALGRAVE | 6592 | 537 | 6055 | 2352 | 0,36 | 88% |
| ROUTLEDGE | 3727 | 280 | 3447 | 1192 | 0,32 | 90% |
| CAMBRIDGE UNIV PRESS | 1544 | 123 | 1421 | 1852 | 1,20 | 82% |
| PRINCETON UNIV PRESS | 1322 | 113 | 1209 | 2768 | 2,09 | 89% |
| UNIV CALIFORNIA PRESS | 1207 | 92 | 1115 | 1244 | 1,03 | 90% |
| UNIV NORTH CAROLINA PRESS | 1144 | 100 | 1044 | 1058 | 0,92 | 89% |
| UNIV PENNSYLVANIA PRESS | 1073 | 108 | 965 | 922 | 0,86 | 85% |
| AUSTRALIAN NATL UNIV | 616 | 38 | 578 | 153 | 0,25 | 89% |
| WALTER DE GRUYTER & CO | 442 | 42 | 400 | 64 | 0,14 | 92% |
| E J BRILL | 385 | 29 | 356 | 14 | 0,04 | 97% |
| SPRINGER | 183 | 13 | 170 | 36 | 0,20 | 93% |
| UTAH STATE UNIV PRESS | 147 | 11 | 136 | 19 | 0,13 | 95% |
| EDINBURGH UNIV PRESS | 139 | 8 | 131 | 4 | 0,03 | 99% |
| PURDUE UNIV PRESS | 123 | 4 | 119 | 1 | 0,01 | 99% |
| UNIV PRESS KENTUCKY | 119 | 8 | 111 | 28 | 0,24 | 96% |
| EDITIONS RODOPI B V | 98 | 9 | 89 | 44 | 0,45 | 84% |
| EDWARD ELGAR PUBLISHING LTD | 95 | 4 | 91 | 32 | 0,34 | 96% |
| FORDHAM UNIV PRESS | 87 | 8 | 79 | 14 | 0,16 | 94% |
| UNIV BRITISH COLUMBIA PRESS | 86 | 7 | 79 | 2 | 0,02 | 98% |
| PICKERING & CHATTO PUBLISHERS | 79 | 9 | 70 | 4 | 0,05 | 95% |
| EMERALD GROUP PUBLISHING LIMITED | 74 | 5 | 69 | 8 | 0,11 | 92% |
| PRESSES UNIV MIRAIL | 74 | 9 | 65 | 5 | 0,07 | 95% |
| BLACKWELL SCIENCE PUBL | 69 | 3 | 66 | 31 | 0,45 | 72% |
| US INST PEACE | 68 | 5 | 63 | 61 | 0,90 | 65% |
| UNIV MINNESOTA PRESS | 65 | 7 | 58 | 18 | 0,28 | 91% |
| WILFRID LAURIER UNIV PRESS | 64 | 4 | 60 | 15 | 0,23 | 95% |
| AMER PHILOSOPHICAL SOC | 56 | 5 | 51 | 2 | 0,04 | 96% |
| MONASH UNIV PUBLISHING | 55 | 4 | 51 | 9 | 0,16 | 93% |
| CENTRAL EUROPEAN UNIV PRESS | 51 | 3 | 48 | 11 | 0,22 | 88% |
| UNIVERSITATSVERLAG WINTER GMBH | 41 | 1 | 40 | 1 | 0,02 | 98% |
| SUOMEN HISTORIALLINEN SEURA | 40 | 4 | 36 | 0 | 0,00 | 100% |
| ATHABASCA UNIV PRESS | 37 | 4 | 33 | 30 | 0,81 | 89% |
| ROYAL HISTORICAL SOC | 32 | 8 | 24 | 2 | 0,06 | 94% |
| INST SOUTHEAST ASIAN STUDIES | 30 | 2 | 28 | 1 | 0,03 | 97% |
| M I T PRESS | 29 | 1 | 28 | 4 | 0,14 | 97% |
| NOTTINGHAM UNIVERSITY PRESS | 29 | 2 | 27 | 1 | 0,03 | 97% |
| INTELLECT LTD | 28 | 2 | 26 | 5 | 0,18 | 96% |
| LIT VERLAG | 26 | 1 | 25 | 0 | 0,00 | 100% |
| NOVA SCIENCE PUBLISHERS, INC | 26 | 4 | 22 | 0 | 0,00 | 100% |
| UNIV PRESS COLORADO | 25 | 2 | 23 | 8 | 0,32 | 92% |
| UNIVERSITY ALASKA PRESS | 24 | 2 | 22 | 5 | 0,21 | 96% |
| IOS PRESS | 23 | 2 | 21 | 0 | 0,00 | 100% |
| CSIRO PUBLISHING | 22 | 2 | 20 | 12 | 0,55 | 91% |
| UNIV ADELAIDE PRESS | 22 | 2 | 20 | 0 | 0,00 | 100% |
| MANEY PUBLISHING | 19 | 3 | 16 | 6 | 0,32 | 89% |
| PSYCHOLOGY PRESS | 18 | 1 | 17 | 4 | 0,22 | 89% |
| WHITE HORSE PRESS | 17 | 2 | 15 | 5 | 0,29 | 88% |
| URBAN LAND INSTITUTE | 14 | 1 | 13 | 6 | 0,43 | 93% |
| PROVERSE HONG KONG | 10 | 1 | 9 | 0 | 0,00 | 100% |
| INFORMATION AGE PUBLISHING-IAP | 9 | 1 | 8 | 1 | 0,11 | 89% |
| SOC NAUTICAL RESEARCH | 9 | 1 | 8 | 4 | 0,44 | 89% |
| TAYLOR & FRANCIS LTD | 9 | 1 | 8 | 1 | 0,11 | 89% |
| BLACKWELL PUBL | 8 | 1 | 7 | 8 | 1,00 | 88% |
| CROATIAN ACAD SCIENCES & ARTS | 7 | 1 | 6 | 0 | 0,00 | 100% |
| WIT PRESS | 6 | 1 | 5 | 0 | 0,00 | 100% |
| EDITURA EXCELSIOR ART | 1 | 1 | | 0 | 0,00 | 100% |
| UNIV WASHINGTON PRESS | 1 | 1 | | 0 | 0,00 | 100% |

## Geography

| | PRODUCTION | | | IMPACT | | |
|---|---|---|---|---|---|---|
| | Total Items | Books | Chap | Total Citations | AvgCit | NonCit |
| SPRINGER | 909 | 71 | 838 | 645 | 0,71 | 76% |
| ROUTLEDGE | 731 | 58 | 673 | 1094 | 1,50 | 75% |
| PALGRAVE | 355 | 31 | 324 | 266 | 0,75 | 84% |
| EDWARD ELGAR PUBLISHING LTD | 134 | 9 | 125 | 183 | 1,37 | 73% |
| UNIV CALIFORNIA PRESS | 81 | 8 | 73 | 215 | 2,65 | 91% |
| AUSTRALIAN NATL UNIV | 60 | 5 | 55 | 73 | 1,22 | 80% |
| CABI PUBLISHING-C A B INT | 56 | 3 | 53 | 37 | 0,66 | 68% |
| PRESSES UNIV MIRAIL | 46 | 2 | 44 | 0 | 0,00 | 100% |
| PRINCETON UNIV PRESS | 35 | 3 | 32 | 37 | 1,06 | 91% |
| UNIV LJUBLJANI, FAK FILOZOFSKA | 35 | 7 | 28 | 0 | 0,00 | 100% |
| ATHABASCA UNIV PRESS | 30 | 2 | 28 | 1 | 0,03 | 97% |
| INST NATL ETUDES DEMOGRAPHIQUES | 20 | 1 | 19 | 0 | 0,00 | 100% |
| UNIV BRITISH COLUMBIA PRESS | 20 | 1 | 19 | 12 | 0,60 | 65% |
| M I T PRESS | 19 | 1 | 18 | 119 | 6,26 | 42% |
| UNIV PENNSYLVANIA PRESS | 18 | 1 | 17 | 9 | 0,50 | 78% |
| MONASH UNIV PUBLISHING | 17 | 1 | 16 | 0 | 0,00 | 100% |
| CSIRO PUBLISHING | 13 | 1 | 12 | 10 | 0,77 | 92% |
| UNIV NORTH CAROLINA PRESS | 13 | 1 | 12 | 0 | 0,00 | 100% |
| BLACKWELL SCIENCE PUBL | 12 | 1 | 11 | 3 | 0,25 | 92% |
| PHYSICA-VERLAG GMBH & CO | 12 | 1 | 11 | 1 | 0,08 | 92% |
| CAMBRIDGE UNIV PRESS | 10 | 1 | 9 | 28 | 2,80 | 90% |
| INST SOUTHEAST ASIAN STUDIES | 10 | 1 | 9 | 0 | 0,00 | 100% |
| SCIENCE PRESS BEIJING | 9 | 1 | 8 | 1 | 0,11 | 89% |
| UNIV MINNESOTA PRESS | 9 | 1 | 8 | 20 | 2,22 | 89% |
| WHITE HORSE PRESS | 7 | 1 | 6 | 0 | 0,00 | 100% |
| UTAH STATE UNIV PRESS | 6 | 1 | 5 | 0 | 0,00 | 100% |
| INT ORGANIZATION MIGRATION | 2 | | 2 | 0 | 0,00 | 100% |
| UNIV WASHINGTON PRESS | 1 | 1 | | 0 | 0,00 | 100% |

Daniel Torres-Salinas, Nicolás Robinson García and Emilio Delgado López-Cózar

# History & Philosophy of Science

| | PRODUCTION | | | IMPACT | | |
|---|---|---|---|---|---|---|
| | Total Items | Books | Chap | Total Citations | AvgCit | NonCit |
| SPRINGER | 2955 | 226 | 2729 | 1312 | 0,44 | 88% |
| PALGRAVE | 496 | 45 | 451 | 220 | 0,44 | 91% |
| ROUTLEDGE | 332 | 24 | 308 | 161 | 0,48 | 83% |
| PRINCETON UNIV PRESS | 290 | 21 | 269 | 124 | 0,43 | 93% |
| CAMBRIDGE UNIV PRESS | 262 | 17 | 245 | 334 | 1,27 | 83% |
| M I T PRESS | 207 | 15 | 192 | 255 | 1,23 | 82% |
| UNIV CALIFORNIA PRESS | 180 | 15 | 165 | 229 | 1,27 | 91% |
| BIRKHAUSER VERLAG AG | 110 | 14 | 96 | 16 | 0,15 | 95% |
| EDITIONS RODOPI B V | 98 | 9 | 89 | 44 | 0,45 | 84% |
| WALTER DE GRUYTER & CO | 95 | 7 | 88 | 21 | 0,22 | 87% |
| CSIRO PUBLISHING | 69 | 4 | 65 | 3 | 0,04 | 97% |
| MATHEMATICAL ASSOC AMERICA | 68 | 2 | 66 | 19 | 0,28 | 84% |
| BLACKWELL SCIENCE PUBL | 53 | 2 | 51 | 68 | 1,28 | 85% |
| WILFRID LAURIER UNIV PRESS | 51 | 3 | 48 | 34 | 0,67 | 78% |
| AUSTRALIAN NATL UNIV | 48 | 4 | 44 | 31 | 0,65 | 79% |
| PAN STANFORD PUBLISHING PTE LTD | 46 | 1 | 45 | 0 | 0,00 | 100% |
| ROYAL SOC CHEMISTRY | 41 | 4 | 37 | 9 | 0,22 | 93% |
| MARY ANN LIEBERT INC | 37 | 2 | 35 | 10 | 0,27 | 92% |
| UNIV NORTH CAROLINA PRESS | 37 | 3 | 34 | 49 | 1,32 | 86% |
| E J BRILL | 36 | 3 | 33 | 0 | 0,00 | 100% |
| AMER INST AERONAUTICS & ASTR | 28 | 2 | 26 | 0 | 0,00 | 100% |
| EDINBURGH UNIV PRESS | 28 | 1 | 27 | 0 | 0,00 | 100% |
| PSYCHOLOGY PRESS | 26 | 2 | 24 | 29 | 1,12 | 77% |
| GEOLOGICAL SOC PUBLISHING HOUSE | 23 | 1 | 22 | 3 | 0,13 | 91% |
| UNIV PENNSYLVANIA PRESS | 22 | 1 | 21 | 21 | 0,95 | 91% |
| AMER ASTRONAUTICAL SOC | 20 | 1 | 19 | 0 | 0,00 | 100% |
| CRC PRESS-TAYLOR & FRANCIS GROUP | 16 | 1 | 15 | 0 | 0,00 | 100% |
| TAYLOR & FRANCIS LTD | 16 | 1 | 15 | 21 | 1,31 | 38% |
| BIRKHAUSER BOSTON | 15 | 1 | 14 | 0 | 0,00 | 100% |
| IOS PRESS | 15 | 1 | 14 | 3 | 0,20 | 80% |
| PURDUE UNIV PRESS | 14 | 1 | 13 | 0 | 0,00 | 100% |
| SCIENCE HISTORY PUBLICATIONS/ USA | 12 | 1 | 11 | 18 | 1,50 | 67% |
| UNIV MINNESOTA PRESS | 12 | 1 | 11 | 33 | 2,75 | 17% |
| A K PETERS, LTD | 11 | 1 | 10 | 2 | 0,18 | 91% |
| JOHNS HOPKINS UNIV PRESS | 10 | 1 | 9 | 1 | 0,10 | 90% |
| ABMS | 9 | 1 | 8 | 0 | 0,00 | 100% |
| INTERVARSITY PRESS | 9 | 1 | 8 | 0 | 0,00 | 100% |
| WORLD SCIENTIFIC PUBL CO PTE LTD | 9 | 1 | 8 | 0 | 0,00 | 100% |
| WELLCOME INST HISTORY MEDICINE | 8 | 1 | 7 | 1 | 0,13 | 88% |
| KARGER | 2 | 1 | 1 | 9 | 4,50 | 50% |
| PROMETHEUS BOOKS | 2 | 2 | | 0 | 0,00 | 100% |
| COLD SPRING HARBOR LABORATORY PRESS | 1 | 1 | | 1 | 1,00 | 0% |

# Information Science & Library Science

| | PRODUCTION | | | IMPACT | | |
|---|---|---|---|---|---|---|
| | Total Items | Books | Chap | Total Citations | AvgCit | NonCit |
| CHANDOS PUBL | 1456 | 125 | 1331 | 502 | 0,34 | 89% |
| IOS PRESS | 760 | 4 | 756 | 202 | 0,27 | 84% |
| SPRINGER | 653 | 44 | 609 | 353 | 0,54 | 81% |
| WALTER DE GRUYTER & CO | 318 | 18 | 300 | 87 | 0,27 | 88% |
| M E SHARPE INC | 252 | 15 | 237 | 175 | 0,69 | 71% |
| BAYWOOD PUBLISHING CO INC | 154 | 13 | 141 | 34 | 0,22 | 85% |
| EMERALD GROUP PUBLISHING LIMITED | 144 | 13 | 131 | 61 | 0,42 | 75% |
| ROUTLEDGE | 101 | 6 | 95 | 14 | 0,14 | 93% |
| PALGRAVE | 100 | 4 | 96 | 7 | 0,07 | 96% |
| M I T PRESS | 47 | 4 | 43 | 34 | 0,72 | 87% |
| WOODHEAD PUBL LTD | 41 | 4 | 37 | 10 | 0,24 | 90% |
| NOVA SCIENCE PUBLISHERS, INC | 28 | 3 | 25 | 0 | 0,00 | 100% |
| CAMBRIDGE UNIV PRESS | 26 | 2 | 24 | 18 | 0,69 | 92% |
| TMC ASSER PRESS | 26 | 1 | 25 | 0 | 0,00 | 100% |
| ELSEVIER | 25 | 2 | 23 | 128 | 5,12 | 92% |
| EDWARD ELGAR PUBLISHING LTD | 23 | 2 | 21 | 31 | 1,35 | 91% |
| CABI PUBLISHING-C A B INT | 21 | 1 | 20 | 50 | 2,38 | 48% |
| WORLD SCIENTIFIC PUBL CO PTE LTD | 18 | 1 | 17 | 8 | 0,44 | 89% |
| UNIV ADELAIDE PRESS | 9 | 1 | 8 | 0 | 0,00 | 100% |
| UTAH STATE UNIV PRESS | 9 | 1 | 8 | 1 | 0,11 | 89% |
| CRC PRESS-TAYLOR & FRANCIS GROUP | 8 | 1 | 7 | 0 | 0,00 | 100% |
| UNIV CALIFORNIA PRESS | 8 | 1 | 7 | 27 | 3,38 | 75% |
| WILFRID LAURIER UNIV PRESS | 8 | 1 | 7 | 3 | 0,38 | 75% |

## Languague & Linguistics

| | PRODUCTION | | | IMPACT | | |
|---|---|---|---|---|---|---|
| | Total Items | Books | Chap | Total Citations | AvgCit | NonCit |
| WALTER DE GRUYTER & CO | 5329 | 301 | 5028 | 2894 | 0,54 | 82% |
| PALGRAVE | 2050 | 169 | 1881 | 1425 | 0,70 | 85% |
| ROUTLEDGE | 1145 | 78 | 1067 | 715 | 0,62 | 86% |
| SPRINGER | 826 | 60 | 766 | 404 | 0,49 | 83% |
| CAMBRIDGE UNIV PRESS | 428 | 29 | 399 | 1029 | 2,40 | 73% |
| M I T PRESS | 292 | 23 | 269 | 551 | 1,89 | 80% |
| PSYCHOLOGY PRESS | 156 | 7 | 149 | 116 | 0,74 | 82% |
| EMERALD GROUP PUBLISHING LIMITED | 136 | 5 | 131 | 71 | 0,52 | 74% |
| NOVA SCIENCE PUBLISHERS, INC | 127 | 9 | 118 | 6 | 0,05 | 95% |
| ELSEVIER | 126 | 11 | 115 | 191 | 1,52 | 67% |
| EDINBURGH UNIV PRESS | 104 | 9 | 95 | 241 | 2,32 | 93% |
| MANEY PUBLISHING | 99 | 7 | 92 | 24 | 0,24 | 90% |
| UTAH STATE UNIV PRESS | 94 | 6 | 88 | 9 | 0,10 | 98% |
| EDITIONS RODOPI B V | 85 | 7 | 78 | 31 | 0,36 | 75% |
| INFORMATION AGE PUBLISHING-IAP | 62 | 7 | 55 | 10 | 0,16 | 89% |
| E J BRILL | 61 | 6 | 55 | 0 | 0,00 | 100% |
| PRINCETON UNIV PRESS | 41 | 3 | 38 | 0 | 0,00 | 100% |
| UNIV CALIFORNIA PRESS | 37 | 3 | 34 | 37 | 1,00 | 92% |
| BLACKWELL SCIENCE PUBL | 36 | 2 | 34 | 13 | 0,36 | 81% |
| LAWRENCE ERLBAUM ASSOC PUBL | 33 | 3 | 30 | 27 | 0,82 | 91% |
| UNIV PENNSYLVANIA PRESS | 32 | 3 | 29 | 28 | 0,88 | 91% |
| GALLAUDET UNIV PR | 30 | 3 | 27 | 18 | 0,60 | 57% |
| METROPOLITNI UNIV PRAHA, OPS | 26 | 1 | 25 | 0 | 0,00 | 100% |
| UNIV PRESS COLORADO | 26 | 1 | 25 | 1 | 0,04 | 96% |
| MONASH UNIV PUBLISHING | 20 | 1 | 19 | 33 | 1,65 | 55% |
| MOUTON DE GRUYTER | 19 | 2 | 17 | 41 | 2,16 | 89% |
| INTELLECT LTD | 14 | 1 | 13 | 10 | 0,71 | 93% |
| POLY PRESS | 12 | 1 | 11 | 0 | 0,00 | 100% |
| WORLD SCIENTIFIC AND ENGI. | 12 | 1 | 11 | 6 | 0,50 | 75% |
| UNIV MANITOBA | 10 | 1 | 9 | 1 | 0,10 | 90% |

Towards a 'Book Publishers Citation Reports'

## Law

| | PRODUCTION | | | IMPACT | | |
|---|---|---|---|---|---|---|
| | Total Items | Books | Chap | Total Citations | AvgCit | NonCit |
| SPRINGER | 2421 | 189 | 2232 | 618 | 0,26 | 90% |
| EDWARD ELGAR PUBLISHING LTD | 1458 | 111 | 1347 | 328 | 0,22 | 89% |
| ROUTLEDGE | 1361 | 105 | 1256 | 217 | 0,16 | 92% |
| WALTER DE GRUYTER & CO | 1180 | 80 | 1100 | 45 | 0,04 | 97% |
| CAMBRIDGE UNIV PRESS | 797 | 63 | 734 | 816 | 1,02 | 77% |
| UNIV PENNSYLVANIA PRESS | 458 | 43 | 415 | 386 | 0,84 | 80% |
| PALGRAVE | 427 | 37 | 390 | 106 | 0,25 | 89% |
| PRINCETON UNIV PRESS | 374 | 34 | 340 | 539 | 1,44 | 82% |
| E J BRILL | 236 | 15 | 221 | 1 | 0,00 | 100% |
| ANNUAL REVIEWS | 132 | 6 | 126 | 593 | 4,49 | 30% |
| EMERALD GROUP PUBLISHING LIMITED | 127 | 15 | 112 | 45 | 0,35 | 81% |
| AUSTRALIAN NATL UNIV | 125 | 8 | 117 | 17 | 0,14 | 90% |
| NOVA SCIENCE PUBLISHERS, INC | 124 | 18 | 106 | 7 | 0,06 | 97% |
| US INST PEACE | 86 | 5 | 81 | 71 | 0,83 | 67% |
| UNIV BRITISH COLUMBIA PRESS | 80 | 7 | 73 | 3 | 0,04 | 98% |
| TAYLOR & FRANCIS LTD | 53 | 3 | 50 | 1 | 0,02 | 98% |
| TMC ASSER PRESS | 43 | 3 | 40 | 1 | 0,02 | 98% |
| UNIV CALIFORNIA PRESS | 43 | 4 | 39 | 47 | 1,09 | 81% |
| UNIV CHICAGO PRESS | 40 | 4 | 36 | 6 | 0,15 | 85% |
| CABI PUBLISHING-C A B INT | 26 | 2 | 24 | 12 | 0,46 | 81% |
| UNIV NORTH CAROLINA PRESS | 26 | 3 | 23 | 23 | 0,88 | 85% |
| BROOKINGS INST | 21 | 2 | 19 | 5 | 0,24 | 86% |
| EDINBURGH UNIV PRESS | 19 | 1 | 18 | 0 | 0,00 | 100% |
| PHYSICA-VERLAG GMBH & CO | 18 | 1 | 17 | 2 | 0,11 | 89% |
| CHANDOS PUBL | 16 | 1 | 15 | 1 | 0,06 | 94% |
| SPON PRESS | 15 | 1 | 14 | 0 | 0,00 | 100% |
| INFORMATION AGE PUBLISHING-IAP | 14 | 1 | 13 | 8 | 0,57 | 50% |
| BLACKWELL SCIENCE PUBL | 13 | 1 | 12 | 5 | 0,38 | 92% |
| MARY ANN LIEBERT INC | 13 | 1 | 12 | 0 | 0,00 | 100% |
| CRC PRESS-TAYLOR & FRANCIS GROUP | 12 | 1 | 11 | 9 | 0,75 | 92% |
| UNIVERSITY ALASKA PRESS | 12 | 1 | 11 | 1 | 0,08 | 92% |
| M I T PRESS | 11 | 1 | 10 | 5 | 0,45 | 91% |
| NOTTINGHAM UNIVERSITY PRESS | 11 | 1 | 10 | 0 | 0,00 | 100% |
| PROVERSE HONG KONG | 10 | 1 | 9 | 0 | 0,00 | 100% |
| CENTRAL EUROPEAN UNIV PRESS | 9 | 1 | 8 | 4 | 0,44 | 89% |
| UNIV ADELAIDE PRESS | 9 | 1 | 8 | 0 | 0,00 | 100% |
| INFORMATION TODAY INC | 4 | 1 | 3 | 0 | 0,00 | 100% |

Daniel Torres-Salinas, Nicolás Robinson García and Emilio Delgado López-Cózar

## Literature

| | Total Items | Books | Chap | Total Citations | AvgCit | NonCit |
|---:|---:|---:|---:|---:|---:|---:|
| PALGRAVE | 4779 | 471 | 4308 | 1224 | 0,26 | 91% |
| WALTER DE GRUYTER & CO | 1452 | 108 | 1344 | 112 | 0,08 | 95% |
| EDITIONS RODOPI B V | 1090 | 61 | 1029 | 117 | 0,11 | 93% |
| CAMBRIDGE UNIV PRESS | 1019 | 89 | 930 | 840 | 0,82 | 77% |
| ROUTLEDGE | 646 | 54 | 592 | 131 | 0,20 | 89% |
| MANEY PUBLISHING | 517 | 54 | 463 | 155 | 0,30 | 88% |
| WILFRID LAURIER UNIV PRESS | 331 | 21 | 310 | 106 | 0,32 | 87% |
| PRINCETON UNIV PRESS | 322 | 33 | 289 | 397 | 1,23 | 86% |
| UNIV PENNSYLVANIA PRESS | 241 | 31 | 210 | 293 | 1,22 | 84% |
| UTAH STATE UNIV PRESS | 213 | 16 | 197 | 78 | 0,37 | 88% |
| PRESSES UNIV MIRAIL | 180 | 15 | 165 | 17 | 0,09 | 95% |
| EDINBURGH UNIV PRESS | 141 | 11 | 130 | 10 | 0,07 | 93% |
| UNIV CALIFORNIA PRESS | 136 | 6 | 130 | 12 | 0,09 | 96% |
| UNIV NORTH CAROLINA PRESS | 98 | 10 | 88 | 81 | 0,83 | 91% |
| FORDHAM UNIV PRESS | 80 | 8 | 72 | 50 | 0,63 | 91% |
| PURDUE UNIV PRESS | 69 | 8 | 61 | 11 | 0,16 | 94% |
| UNIV PITTSBURGH PRESS | 60 | 6 | 54 | 3 | 0,05 | 95% |
| E J BRILL | 42 | 3 | 39 | 0 | 0,00 | 100% |
| ASHGATE PUBLISHING LTD | 37 | 2 | 35 | 14 | 0,38 | 76% |
| UNIV MINNESOTA PRESS | 31 | 3 | 28 | 9 | 0,29 | 94% |
| INNSBRUCK UNIV PRESS | 27 | 1 | 26 | 0 | 0,00 | 100% |
| PICKERING & CHATTO PUBLISHERS | 27 | 2 | 25 | 2 | 0,07 | 93% |
| AUSTRALIAN NATL UNIV | 26 | 2 | 24 | 0 | 0,00 | 100% |
| DUKE UNIV PRESS | 23 | 1 | 22 | 0 | 0,00 | 100% |
| CHANDOS PUBL | 17 | 1 | 16 | 18 | 1,06 | 76% |
| INTELLECT LTD | 13 | 1 | 12 | 3 | 0,23 | 77% |
| UNIV NACIONAL AUTONOMA MEXICO | 10 | 1 | 9 | 2 | 0,20 | 90% |
| UNIV BRITISH COLUMBIA PRESS | 9 | 1 | 8 | 0 | 0,00 | 100% |
| GALLAUDET UNIV PR | 8 | 1 | 7 | 3 | 0,38 | 88% |
| SPRINGER | 6 | 1 | 5 | 0 | 0,00 | 100% |
| LEGENDA-MODERN | 3 | 3 | | 0 | 0,00 | 100% |
| RENAISSANCE INST | 1 | 1 | | 1 | 1,00 | 0% |

## Managment

| | PRODUCTION | | | IMPACT | | |
|---|---|---|---|---|---|---|
| | Total Items | Books | Chap | Total Citations | AvgCit | NonCit |
| EDWARD ELGAR PUBLISHING LTD | 1509 | 102 | 1407 | 765 | 0,51 | 82% |
| ROUTLEDGE | 1134 | 77 | 1057 | 636 | 0,56 | 83% |
| PALGRAVE | 1086 | 70 | 1016 | 96 | 0,09 | 96% |
| SPRINGER | 1040 | 88 | 952 | 304 | 0,29 | 88% |
| EMERALD GROUP PUBLISHING LIMITED | 754 | 56 | 698 | 920 | 1,22 | 64% |
| INFORMATION AGE PUBLISHING-IAP | 390 | 30 | 360 | 164 | 0,42 | 83% |
| CHANDOS PUBL | 221 | 16 | 205 | 51 | 0,23 | 94% |
| PHYSICA-VERLAG GMBH & CO | 209 | 19 | 190 | 79 | 0,38 | 87% |
| CAMBRIDGE UNIV PRESS | 191 | 11 | 180 | 137 | 0,72 | 81% |
| WORLD SCIENTIFIC PUBL CO PTE LTD | 93 | 4 | 89 | 8 | 0,09 | 94% |
| PRINCETON UNIV PRESS | 89 | 6 | 83 | 362 | 4,07 | 90% |
| IOS PRESS | 82 | 7 | 75 | 18 | 0,22 | 87% |
| TAYLOR & FRANCIS LTD | 80 | 5 | 75 | 15 | 0,19 | 93% |
| PSYCHOLOGY PRESS | 70 | 4 | 66 | 200 | 2,86 | 51% |
| JAI-ELSEVIER SCIENCE INC | 67 | 6 | 61 | 349 | 5,21 | 61% |
| CRC PRESS-TAYLOR & FRANCIS GROUP | 62 | 4 | 58 | 4 | 0,06 | 94% |
| JAI-ELSEVIER LTD | 58 | 4 | 54 | 14 | 0,24 | 83% |
| ELSEVIER | 55 | 4 | 51 | 90 | 1,64 | 55% |
| AUERBACH PUBLICATIONS | 50 | 4 | 46 | 10 | 0,20 | 92% |
| JAI PRESS INC | 49 | 3 | 46 | 14 | 0,29 | 84% |
| IGI GLOBAL | 44 | 3 | 41 | 9 | 0,20 | 95% |
| EDINBURGH UNIV PRESS | 31 | 3 | 28 | 1 | 0,03 | 97% |
| APRESS | 27 | 1 | 26 | 0 | 0,00 | 100% |
| NOVA SCIENCE PUBLISHERS, INC | 25 | 2 | 23 | 0 | 0,00 | 100% |
| UNIV PENNSYLVANIA PRESS | 25 | 2 | 23 | 109 | 4,36 | 36% |
| WOODHEAD PUBL LTD | 22 | 2 | 20 | 0 | 0,00 | 100% |
| M I T PRESS | 18 | 1 | 17 | 27 | 1,50 | 61% |
| AMER INST AERONAUTICS & ASTR. | 16 | 1 | 15 | 0 | 0,00 | 100% |
| PURDUE UNIV PRESS | 14 | 1 | 13 | 6 | 0,43 | 79% |
| SPON PRESS | 14 | 1 | 13 | 0 | 0,00 | 100% |
| BLACKWELL SCIENCE PUBL | 13 | 1 | 12 | 0 | 0,00 | 100% |
| CA PRESS | 13 | 1 | 12 | 0 | 0,00 | 100% |
| PRODUCTIVITY PRESS | 13 | 1 | 12 | 0 | 0,00 | 100% |
| WALTER DE GRUYTER & CO | 13 | 1 | 12 | 1 | 0,08 | 92% |
| COLD SPRING HARBOR LABORATORY | 10 | 1 | 9 | 0 | 0,00 | 100% |
| VS VERLAG SOZIALWISSENSCHAFTEN- | 10 | 1 | 9 | 0 | 0,00 | 100% |

Daniel Torres-Salinas, Nicolás Robinson García and Emilio Delgado López-Cózar

# Philosophy & Ethics

| | PRODUCTION | | | IMPACT | | |
|---|---|---|---|---|---|---|
| | Total Items | Books | Chap | Total Citations | AvgCit | NonCit |
| SPRINGER | 3210 | 228 | 2982 | 754 | 0,23 | 89% |
| PALGRAVE | 2521 | 208 | 2313 | 1069 | 0,42 | 89% |
| ROUTLEDGE | 1923 | 141 | 1782 | 650 | 0,34 | 89% |
| WALTER DE GRUYTER & CO | 1372 | 95 | 1277 | 259 | 0,19 | 90% |
| CAMBRIDGE UNIV PRESS | 1230 | 93 | 1137 | 1495 | 1,22 | 76% |
| PRINCETON UNIV PRESS | 894 | 79 | 815 | 1453 | 1,63 | 86% |
| M I T PRESS | 190 | 16 | 174 | 459 | 2,42 | 73% |
| INFORMATION AGE PUBLISHING-IAP | 174 | 12 | 162 | 88 | 0,51 | 79% |
| EDITIONS RODOPI B V | 148 | 9 | 139 | 39 | 0,26 | 80% |
| EDINBURGH UNIV PRESS | 128 | 12 | 116 | 4 | 0,03 | 98% |
| EDWARD ELGAR PUBLISHING LTD | 67 | 4 | 63 | 78 | 1,16 | 63% |
| BLACKWELL SCIENCE PUBL | 59 | 4 | 55 | 40 | 0,68 | 76% |
| UNIV CALIFORNIA PRESS | 51 | 4 | 47 | 121 | 2,37 | 82% |
| FORDHAM UNIV PRESS | 49 | 5 | 44 | 282 | 5,76 | 90% |
| WILEY-BLACKWELL | 37 | 2 | 35 | 6 | 0,16 | 86% |
| NOVA SCIENCE PUBLISHERS, INC | 36 | 3 | 33 | 16 | 0,44 | 92% |
| IL POLIGRAFO | 35 | 3 | 32 | 4 | 0,11 | 91% |
| KLUWER ACADEMIC PUBLISHERS | 35 | 2 | 33 | 8 | 0,23 | 86% |
| AUSTRALIAN NATL UNIV | 32 | 2 | 30 | 12 | 0,38 | 72% |
| E J BRILL | 27 | 2 | 25 | 1 | 0,04 | 96% |
| PURDUE UNIV PRESS | 26 | 3 | 23 | 3 | 0,12 | 92% |
| INDEPENDENT INST | 22 | 1 | 21 | 5 | 0,23 | 86% |
| PRESSES UNIV MIRAIL | 20 | 2 | 18 | 1 | 0,05 | 95% |
| GLOBAL ACADEMIC PUBLISHING | 19 | 3 | 16 | 1 | 0,05 | 95% |
| UNIV PENNSYLVANIA PRESS | 18 | 3 | 15 | 11 | 0,61 | 89% |
| TAYLOR & FRANCIS LTD | 17 | 1 | 16 | 0 | 0,00 | 100% |
| GALLAUDET UNIV PR | 13 | 1 | 12 | 15 | 1,15 | 54% |
| ATHABASCA UNIV PRESS | 8 | 1 | 7 | 0 | 0,00 | 100% |
| CHANDOS PUBL | 8 | 1 | 7 | 0 | 0,00 | 100% |
| UNIV MINNESOTA PRESS | 8 | 1 | 7 | 12 | 1,50 | 75% |
| INTELLECT LTD | 7 | 1 | 6 | 0 | 0,00 | 100% |
| MANEY PUBLISHING | 7 | 1 | 6 | 0 | 0,00 | 100% |
| CUADERNOS DE ANUARIO FILOSOFICO | 1 | 1 | | 1 | 1,00 | 0% |

# Political Science & International Relations

| | PRODUCTION | | | IMPACT | | |
|---|---|---|---|---|---|---|
| | Total Items | Books | Chap | Total Citations | AvgCit | NonCit |
| PALGRAVE | 12306 | 1045 | 11261 | 7443 | 0,60 | 84% |
| ROUTLEDGE | 9156 | 758 | 8398 | 4431 | 0,48 | 86% |
| PRINCETON UNIV PRESS | 2230 | 204 | 2026 | 6201 | 2,78 | 85% |
| CAMBRIDGE UNIV PRESS | 1241 | 103 | 1138 | 3036 | 2,45 | 77% |
| SPRINGER | 1153 | 110 | 1043 | 470 | 0,41 | 85% |
| EDWARD ELGAR PUBLISHING LTD | 1046 | 84 | 962 | 857 | 0,82 | 78% |
| UNIV PENNSYLVANIA PRESS | 659 | 66 | 593 | 897 | 1,36 | 80% |
| AUSTRALIAN NATL UNIV | 497 | 35 | 462 | 110 | 0,22 | 88% |
| EMERALD GROUP PUBLISHING LIMITED | 456 | 33 | 423 | 88 | 0,19 | 85% |
| US INST PEACE | 434 | 56 | 378 | 399 | 0,92 | 81% |
| NOVA SCIENCE PUBLISHERS, INC | 365 | 50 | 315 | 24 | 0,07 | 97% |
| UNIV CALIFORNIA PRESS | 325 | 23 | 302 | 223 | 0,69 | 90% |
| WILFRID LAURIER UNIV PRESS | 222 | 14 | 208 | 111 | 0,50 | 72% |
| M I T PRESS | 194 | 15 | 179 | 423 | 2,18 | 72% |
| ANNUAL REVIEWS | 133 | 6 | 127 | 1553 | 11,68 | 21% |
| UNIV BRITISH COLUMBIA PRESS | 104 | 9 | 95 | 49 | 0,47 | 79% |
| E J BRILL | 101 | 8 | 93 | 8 | 0,08 | 95% |
| TAYLOR & FRANCIS LTD | 97 | 6 | 91 | 14 | 0,14 | 96% |
| BROOKINGS INST | 94 | 8 | 86 | 24 | 0,26 | 87% |
| EDINBURGH UNIV PRESS | 92 | 7 | 85 | 5 | 0,05 | 99% |
| UNIV NORTH CAROLINA PRESS | 87 | 8 | 79 | 45 | 0,52 | 92% |
| UNIV ADELAIDE PRESS | 86 | 4 | 82 | 0 | 0,00 | 100% |
| WALTER DE GRUYTER & CO | 86 | 8 | 78 | 2 | 0,02 | 98% |
| IOS PRESS | 67 | 3 | 64 | 12 | 0,18 | 91% |
| VS VERLAG SOZIALWISSENSCHAFTEN | 60 | 5 | 55 | 18 | 0,30 | 85% |
| INST SOUTHEAST ASIAN STUDIES | 54 | 16 | 38 | 26 | 0,48 | 87% |
| EAST-WEST CENTER | 43 | 34 | 9 | 134 | 3,12 | 35% |
| WORLD SCIENTIFIC PUBL CO PTE LTD | 43 | 3 | 40 | 12 | 0,28 | 84% |
| UNIVERSITATSVERLAG WINTER GMBH | 41 | 1 | 40 | 1 | 0,02 | 98% |
| UNIV MINNESOTA PRESS | 40 | 4 | 36 | 25 | 0,63 | 85% |
| INFORMATION AGE PUBLISHING-IAP | 36 | 3 | 33 | 1 | 0,03 | 97% |
| INTELLECT LTD | 31 | 2 | 29 | 41 | 1,32 | 71% |
| CENTRAL EUROPEAN UNIV PRESS | 30 | 2 | 28 | 59 | 1,97 | 90% |
| PSYCHOLOGY PRESS | 24 | 1 | 23 | 0 | 0,00 | 100% |
| UNIVERSITY ALASKA PRESS | 21 | 1 | 20 | 0 | 0,00 | 100% |
| PHYSICA-VERLAG GMBH & CO | 18 | 1 | 17 | 2 | 0,11 | 89% |
| CENTRUM PRO STUDIUM DEMOKRACIE | 17 | 1 | 16 | 1 | 0,06 | 94% |
| WESTDEUTSCHER VERLAG GMBH | 15 | 2 | 13 | 2 | 0,13 | 93% |
| JOHN WILEY & SONS | 14 | 1 | 13 | 1 | 0,07 | 93% |
| ROWMAN & LITTLEFIELD PUBL INC | 13 | 1 | 12 | 74 | 5,69 | 23% |
| FORDHAM UNIV PRESS | 10 | 1 | 9 | 8 | 0,80 | 90% |
| GEORGETOWN UNIV PRESS | 9 | 1 | 8 | 2 | 0,22 | 89% |
| BLACKWELL PUBL | 8 | 1 | 7 | 8 | 1,00 | 88% |
| MATFYZPRESS PUBL HOUSE | 7 | 1 | 6 | 1 | 0,14 | 86% |
| PRESSES UNIV MIRAIL | 7 | 1 | 6 | 1 | 0,14 | 86% |
| TMC ASSER PRESS | 7 | 1 | 6 | 0 | 0,00 | 100% |
| WORLD HEALTH ORGANIZATION | 6 | 1 | 5 | 9 | 1,50 | 83% |
| INFORMATION TODAY INC | 4 | 1 | 3 | 0 | 0,00 | 100% |
| UNITED STATES INSTITUTE OF PEACE | 1 | 1 | | 0 | 0,00 | 100% |

Daniel Torres-Salinas, Nicolás Robinson García and Emilio Delgado López-Cózar

## Religion

| | PRODUCTION | | | IMPACT | | |
|---|---|---|---|---|---|---|
| | Total Items | Books | Chap | Total Citations | AvgCit | NonCit |
| PALGRAVE | 1622 | 138 | 1484 | 490 | 0,30 | 92% |
| WALTER DE GRUYTER & CO | 1417 | 128 | 1289 | 190 | 0,13 | 94% |
| ROUTLEDGE | 1151 | 94 | 1057 | 241 | 0,21 | 92% |
| PRINCETON UNIV PRESS | 698 | 57 | 641 | 916 | 1,31 | 86% |
| CAMBRIDGE UNIV PRESS | 565 | 49 | 516 | 616 | 1,09 | 80% |
| SPRINGER | 562 | 39 | 523 | 62 | 0,11 | 93% |
| E J BRILL | 544 | 39 | 505 | 16 | 0,03 | 98% |
| UNIV CALIFORNIA PRESS | 543 | 48 | 495 | 586 | 1,08 | 91% |
| UNIV PENNSYLVANIA PRESS | 411 | 43 | 368 | 304 | 0,74 | 85% |
| WILFRID LAURIER UNIV PRESS | 207 | 15 | 192 | 48 | 0,23 | 90% |
| UNIV NORTH CAROLINA PRESS | 178 | 17 | 161 | 171 | 0,96 | 82% |
| BLACKWELL SCIENCE PUBL | 134 | 6 | 128 | 56 | 0,42 | 81% |
| FORDHAM UNIV PRESS | 121 | 7 | 114 | 8 | 0,07 | 95% |
| AUSTRALIAN NATL UNIV | 108 | 9 | 99 | 26 | 0,24 | 90% |
| PURDUE UNIV PRESS | 79 | 3 | 76 | 1 | 0,01 | 99% |
| M I T PRESS | 62 | 1 | 61 | 5 | 0,08 | 98% |
| INFORMATION AGE PUBLISHING-IAP | 47 | 4 | 43 | 1 | 0,02 | 98% |
| EDWARD ELGAR PUBLISHING LTD | 36 | 3 | 33 | 1 | 0,03 | 97% |
| UTAH STATE UNIV PRESS | 32 | 5 | 27 | 3 | 0,09 | 94% |
| NOVA SCIENCE PUBLISHERS, INC | 30 | 4 | 26 | 7 | 0,23 | 83% |
| UNIV BRITISH COLUMBIA PRESS | 20 | 2 | 18 | 5 | 0,25 | 95% |
| CABI PUBLISHING-C A B INT | 18 | 1 | 17 | 16 | 0,89 | 61% |
| EDINBURGH UNIV PRESS | 18 | 2 | 16 | 6 | 0,33 | 94% |
| CENTRUM PRO STUDIUM DEMOKRACIE | 17 | 1 | 16 | 1 | 0,06 | 94% |
| WAGENINGEN ACADEMIC PUBLISHERS | 17 | 1 | 16 | 1 | 0,06 | 94% |
| ROYAL COLL PSYCHIATRISTS | 15 | 1 | 14 | 10 | 0,67 | 73% |
| TAYLOR & FRANCIS LTD | 13 | 1 | 12 | 0 | 0,00 | 100% |
| INTERVARSITY PRESS | 9 | 1 | 8 | 0 | 0,00 | 100% |
| MANEY PUBLISHING | 9 | 1 | 8 | 5 | 0,56 | 89% |
| US INST PEACE | 1 | 1 | | 3 | 3,00 | 0% |

## SOCIOLOGY

| | PRODUCTION | | | IMPACT | | |
|---|---|---|---|---|---|---|
| | Total Items | Books | Chap | Total Citations | AvgCit | NonCit |
| PALGRAVE | 2454 | 209 | 2245 | 1587 | 0,65 | 84% |
| ROUTLEDGE | 1702 | 140 | 1562 | 1251 | 0,74 | 89% |
| SPRINGER | 940 | 63 | 877 | 460 | 0,49 | 78% |
| EMERALD GROUP PUBLISHING LIMITED | 926 | 66 | 860 | 723 | 0,78 | 69% |
| UNIV CALIFORNIA PRESS | 667 | 53 | 614 | 1366 | 2,05 | 90% |
| PRINCETON UNIV PRESS | 460 | 38 | 422 | 2014 | 4,38 | 79% |
| ANNUAL REVIEWS | 303 | 13 | 290 | 3048 | 10,06 | 23% |
| CAMBRIDGE UNIV PRESS | 225 | 18 | 207 | 588 | 2,61 | 72% |
| CABI PUBLISHING-C A B INT | 204 | 12 | 192 | 349 | 1,71 | 53% |
| WILEY-BLACKWELL | 154 | 12 | 142 | 829 | 5,38 | 41% |
| WALTER DE GRUYTER & CO | 136 | 10 | 126 | 55 | 0,40 | 83% |
| JAI PRESS INC | 107 | 9 | 98 | 219 | 2,05 | 41% |
| AUSTRALIAN NATL UNIV | 105 | 8 | 97 | 22 | 0,21 | 84% |
| NOVA SCIENCE PUBLISHERS, INC | 105 | 6 | 99 | 8 | 0,08 | 93% |
| UNIV MINNESOTA PRESS | 73 | 6 | 67 | 35 | 0,48 | 84% |
| EDWARD ELGAR PUBLISHING LTD | 66 | 6 | 60 | 60 | 0,91 | 67% |
| PRESSES UNIV MIRAIL | 58 | 6 | 52 | 5 | 0,09 | 97% |
| E J BRILL | 48 | 5 | 43 | 7 | 0,15 | 96% |
| M I T PRESS | 37 | 3 | 34 | 289 | 7,81 | 46% |
| UNIV PENNSYLVANIA PRESS | 33 | 3 | 30 | 221 | 6,70 | 73% |
| TAYLOR & FRANCIS LTD | 32 | 3 | 29 | 4 | 0,13 | 94% |
| UNIV NORTH CAROLINA PRESS | 32 | 3 | 29 | 17 | 0,53 | 84% |
| BLACKWELL SCIENCE PUBL | 30 | 1 | 29 | 18 | 0,60 | 63% |
| PURDUE UNIV PRESS | 30 | 2 | 28 | 26 | 0,87 | 60% |
| WILFRID LAURIER UNIV PRESS | 28 | 2 | 26 | 29 | 1,04 | 43% |
| LIT VERLAG | 26 | 1 | 25 | 0 | 0,00 | 100% |
| UNIV BRITISH COLUMBIA PRESS | 23 | 2 | 21 | 13 | 0,57 | 74% |
| EDINBURGH UNIV PRESS | 19 | 2 | 17 | 130 | 6,84 | 79% |
| VS VERLAG SOZIALWISSENSCHAFTEN | 16 | 1 | 15 | 2 | 0,13 | 94% |
| WORLD HEALTH ORGANIZATION | 15 | 1 | 14 | 49 | 3,27 | 20% |
| US INST PEACE | 10 | 1 | 9 | 34 | 3,40 | 90% |
| JAI-ELSEVIER SCI BV | 8 | 1 | 7 | 6 | 0,75 | 63% |
| INTELLECT LTD | 7 | 1 | 6 | 0 | 0,00 | 100% |
| CRC PRESS-TAYLOR & FRANCIS GROUP | 1 | | 1 | 0 | 0,00 | 100% |

Daniel Torres-Salinas, Nicolás Robinson García and Emilio Delgado López-Cózar

## Appendix 2. Disciplines and Web of Science subject categories

We show in this appendix how disciplines used in this study are configured according Web of Science Categories.

| Disciplines used in this study | Web of Science Category assigned |
|---|---|
| Anthropology | Anthropology |
| Archaeology | Archaeology |
| Area & Cultural Studies | Cultural Studies |
| | Social Issues |
| | Area Studies |
| | Asian Studies |
| Arts | Art |
| Communication | Film, Radio, Television |
| | Communication |
| Economics & Bussiness | Industrial Relations & Labor |
| | Business, Finance |
| | Business |
| | Economics |
| Education | Education & Educational Research |
| | Education, Scientific Disciplines |
| | Education, Special |
| | Psychology, Educational |
| Geography | Geography |
| | Demography |
| History | History |
| History & Philosophy of Science | History & Philosophy Of Science |
| Information Science & Library Science | Information Science & Library Science |
| Languague & Linguistics | Language & Linguistics |
| | Linguistics |
| Law | Law |
| Literature | Literature, American |
| | Poetry |
| | Literature, Slavic |
| | Literature, Romance |
| | Literature, British Isles |
| | Literature, African, Australian, Canadian |
| | Literature |
| | Literature, German, Dutch, Scandinavian |
| Managment | Management |
| Philosophy & Ethics | Ethics |
| | Philosophy |
| Political Science & International Relations | International Relations |
| | Political Science |
| Religion | Religion |
| Sociology | Sociology |

EC³ Working Papers